\documentclass[aps,prl,letterpaper,twocolumn,superscriptaddress]{revtex4-2}
\usepackage{graphicx} 
\usepackage{float}
\usepackage{tikz}
\usepackage{amsmath}
\usepackage{amssymb}
\usepackage{physics}

\newcommand{\emptycircle}{
  \begin{tikzpicture}[baseline=-0.6ex]
    \draw (0,0) circle (0.1);
  \end{tikzpicture}
}
\newcommand{\fullcircle}{
  \begin{tikzpicture}[baseline=-0.6ex]
     \filldraw[fill=black] (0,0) circle (0.1);
  \end{tikzpicture}
}

\begin{document}
\title{Infinite randomness criticality and localization of the floating phase in arrays of Rydberg atoms trapped with non-perfect tweezers}


\author{Jose Soto-Garcia}
\affiliation{Kavli Institute of Nanoscience, Delft University of Technology, Lorentzweg 1, 2628 CJ Delft, The Netherlands}
\author{Natalia Chepiga}
\affiliation{Kavli Institute of Nanoscience, Delft University of Technology, Lorentzweg 1, 2628 CJ Delft, The Netherlands}

\begin{abstract}

Chains of Rydberg atoms have emerged as a powerful platform for exploring low-dimensional quantum physics. This success originates from the precise control of lattice geometries provided by optical tweezers, which allows access to a wide range of synthetic quantum phases. Experiments on one-dimensional arrays have stimulated tremendous progress in understanding quantum phase transitions into crystalline phases. However, the finite width of tweezers introduces small variations in interatomic distances, leading to quenched disorder in the interactions. In this letter, we numerically study how such disorder alters the nature of two critical regimes observed in experiments. Firstly, following experimental protocols, we analyze Kibble–Zurek dynamics and find a crossover from the clean Ising transition to the infinite-randomness fixed point as system size and disorder strength increase. Secondly, we show that the floating phase---an incommensurate Luttinger liquid phase emerging at stronger interactions---is localized by the disorder, yet preserves short-range incommensurate correlations with the same leading wave vector. Our results clearly reveal an additional conceptual challenge in understanding critical phenomena using Rydberg-based quantum simulators.

\end{abstract}

\maketitle
{\bf Introduction.} Rydberg atom arrays have emerged as a versatile and highly controllable platform for exploring quantum many-body physics \cite{bernien2017probing, keesling2019quantum, ebadi2021quantum, scholl2021quantum}. By trapping individual atoms in optical tweezers and exciting them to a Rydberg state, experiments have realized a wide variety of interacting systems with remarkable precision \cite{zhang2025probing,labuhn2016tunable, de2019observation, verresen2019stable, celi2020emerging,  chen2023continuous}. Despite the apparent simplicity of the underlying models, these systems have revealed many exotic phases and quantum phase transitions~\cite{chepiga2019floating,chepiga2021kibble,rader2019floating,PhysRevA.98.023614,chepiga2024tunable, surace2020lattice, li2024uncovering, soto2025numerical, wang2025tricritical, brodoloni2025spin, zhang2025probing, bock2025parton,eck2023critical}. In particular, Rydberg atoms provide an experimentally accessible platform for commensurate–incommensurate (C–IC) transitions \cite{bak1982commensurate}, which arise from the competition between incompatible lattice orders \cite{bak1982commensurate, huse1982domain, huse1984commensurate}. This competition generates a rich phase structure, including critical floating phases and unconventional chiral transitions \cite{huse1982domain, chepiga2019floating, chepiga2021kibble, rader2019floating, PhysRevA.98.023614}, where competing interactions are responsible for incommensurate correlations.

Beyond equilibrium physics, Rydberg chains have been instrumental in probing many non-equilibrium quantum phenomena. Notably, they have served as a testing ground  for studying quantum many-body scars \cite{bernien2017probing, turner2018weak, serbyn2021quantum, moudgalya2022quantum}---atypical eigenstates that are believed to violate the eigenstate thermalization hypothesis---and for exploring the quantum version of the celebrated Kibble-Zurek (KZ) mechanism \cite{zurek2005dynamics, dziarmaga2010dynamics, keesling2019quantum, wang2025tricritical, soto2024resolving}, which describes the universal scaling of defects generated when a system is driven with a constant rate through a continuous phase transition. This theory predicts that when quenching with a slow rate through a second order phase transition, the density of kinks scales as a power law with the sweep rate:
\begin{equation}
    n_k \sim s^{\mu},\quad \textnormal{with} \quad \mu = \frac{\nu}{1 + z\nu},
\end{equation}
where $\mu$ is the KZ critical exponent, which relates the critical exponent $\nu$ and the dynamical critical exponent $z$.

A key advantage of these platforms is the control over atomic positions and interaction strengths. In practice, however, fluctuations of atomic positions inside each tweezer---especially along the axis perpendicular to the plane of the lattice---introduce an intrinsic level of spatial disorder (see, for instance, the appendix of Ref.\cite{bernien2017probing}). While typically small, such distortions can qualitatively alter critical behavior \cite{wang2025lattice, prodius2025interplay}.

The paradigmatic example demonstrating the significant effect of disorder on quantum phase transitions in one dimension is infinite randomness criticality. It appears as a fixed point in the transverse-field Ising model in the presence of disorder---random distribution either in the interaction or in the transverse field or in both \cite{fisher1992random, fisher1995critical}. It was shown that such disorder forms a relevant perturbation: even an infinitesimal width in the random distribution of coupling constants brings the system into a fundamentally different universality class in the thermodynamic limit. This regime is governed by extremely anisotropic scaling, with two distinct correlation length critical exponents $\nu_\mathrm{typ} = 1$ for typical and $\nu_\mathrm{av} = 2$ for average correlations, and an infinite dynamical critical exponent, $z \to \infty$~\cite{fisher1992random, fisher1995critical, young1996numerical}, forming a striking contrast to the clean Ising transition with a single $\nu=1$ and $z=1$. 
The crossover between these two universality classes---clean Ising and infinite randomness--- is governed by system size, disorder strength, and proximity to criticality~\cite{PhysRevB.70.054430,fisher1995critical, dziarmaga2006dynamics, caneva2007adiabatic}. 

These changes in universality class have profound consequences for the KZ mechanism. In clean Ising systems, the density of topological defects (kinks) scales with the sweep rate of the control parameter according to a universal power law $\mu = 0.5$~\cite{zurek2005dynamics}. In contrast, in the infinite-randomness criticality, this scaling is dramatically modified: the defect density no longer follows a power law, but exhibits a logarithmic dependence on the sweep rate~\cite{dziarmaga2006dynamics, caneva2007adiabatic}. Interestingly, this anomalous scaling behavior only emerges at sufficiently slow sweeps, while faster ramps  exhibit a crossover resembling the clean Ising regime~\cite{dziarmaga2006dynamics, caneva2007adiabatic}.

In this Letter, we study how disorder in the Rydberg interactions impacts two prominent physical phenomena recently highlighted in the literature: {\it (i)} the KZ probe of the Ising transition into the period-2 phase; and {\it (ii)} the equilibrium properties of the floating phase. Our focus on the Ising transition is motivated by the fact that, to date, it is the only case where the effects of disorder on both equilibrium and KZ scaling are theoretically well-understood~\cite{igloi2005strong}, though the role of longer-range interactions is still debated~\cite{fisher1999phase, juhasz2014random, li2016many}. By contrast, the effect of disorder on the chiral transitions appearing at the boundary of the period-3 and 4 lobes~\cite{huse1982domain,huse1984commensurate,chepiga2019floating,chepiga2021kibble,rader2019floating,maceira2022conformal}, to the best of our knowledge, remains unexplored. While we do not address this problem directly, we argue that disorder can explain systematic discrepancies between numerical simulations and experimental results~\cite{keesling2019quantum}. Finally, we show that disorder has a dramatic effect on the floating phase: it destroys quantum criticality while preserving short-range incommensurate correlations with the same leading wave vector.

{\bf The model.} 
We consider a finite one-dimensional chain of Rydberg atoms, described in terms of hard-core bosons with a disordered interaction potential:

\begin{equation}
\frac{H}{\hbar} = \frac{\Omega}{2} \sum_i \left( d_i + d_i^\dagger \right) - \Delta \sum_i n_i + \sum_{i<j} V_{i} \frac{n_i n_j}{(i-j)^6},
\end{equation}
where $d_i^\dagger$ ($d_i$) creates (annihilates) a Rydberg excitation at site $i$, and $n_i = d_i^\dagger d_i$ is the corresponding number operator. The hard-core constraint enforces a single excitation per site, i.e. $d_i^\dagger \ket{1}_i = 0$.  $\Omega$ is the Rabi frequency and $\Delta$ is the laser detuning of the coherent laser with respect to the resonant frequency.  The blockade effect prevents simultaneous excitation of atoms within a certain distance, characterized by the blockade radius $R_b/a = (V_0 / \Omega)^{1/6}$, with $a$ being the interatomic distance.

We assume that tweezers form an evenly spaced lattice, while deviations in the atomic positions are introduced through a  disorder in the interaction strength. Thus, the potential $V_i$ is a quenched random variable with a uniform box distribution in the interval $[1 - \delta V, 1 + \delta V]$.

Based on the typical uncertainties on the position of the atoms~\cite{bernien2017probing}, we study four values of disorder strength $\delta V = 0.15$, $0.25$, $0.4$ and $0.5$. Further details can be found in the End Matter. We also include the clean case $\delta V = 0$ for the reference.

Let us first examine how the direct Ising transition in arrays of Rydberg atoms is influenced by uncertainties in the atomic positions, closely following the experimental protocol of Ref.~\cite{bernien2017probing} within the framework of the quantum Kibble–Zurek (KZ) mechanism. Specifically, we investigate the scaling of the kink density $n_k$—the density of domain walls formed after a slow quench across the Ising transition from the period-2 ordered phase—as a function of disorder strength $\delta V$ and system size $L$. A microscopic definition of the kink operator employed in this study is provided in the End Matter.

\begin{figure}[t!]
    \includegraphics[width=0.45\textwidth]{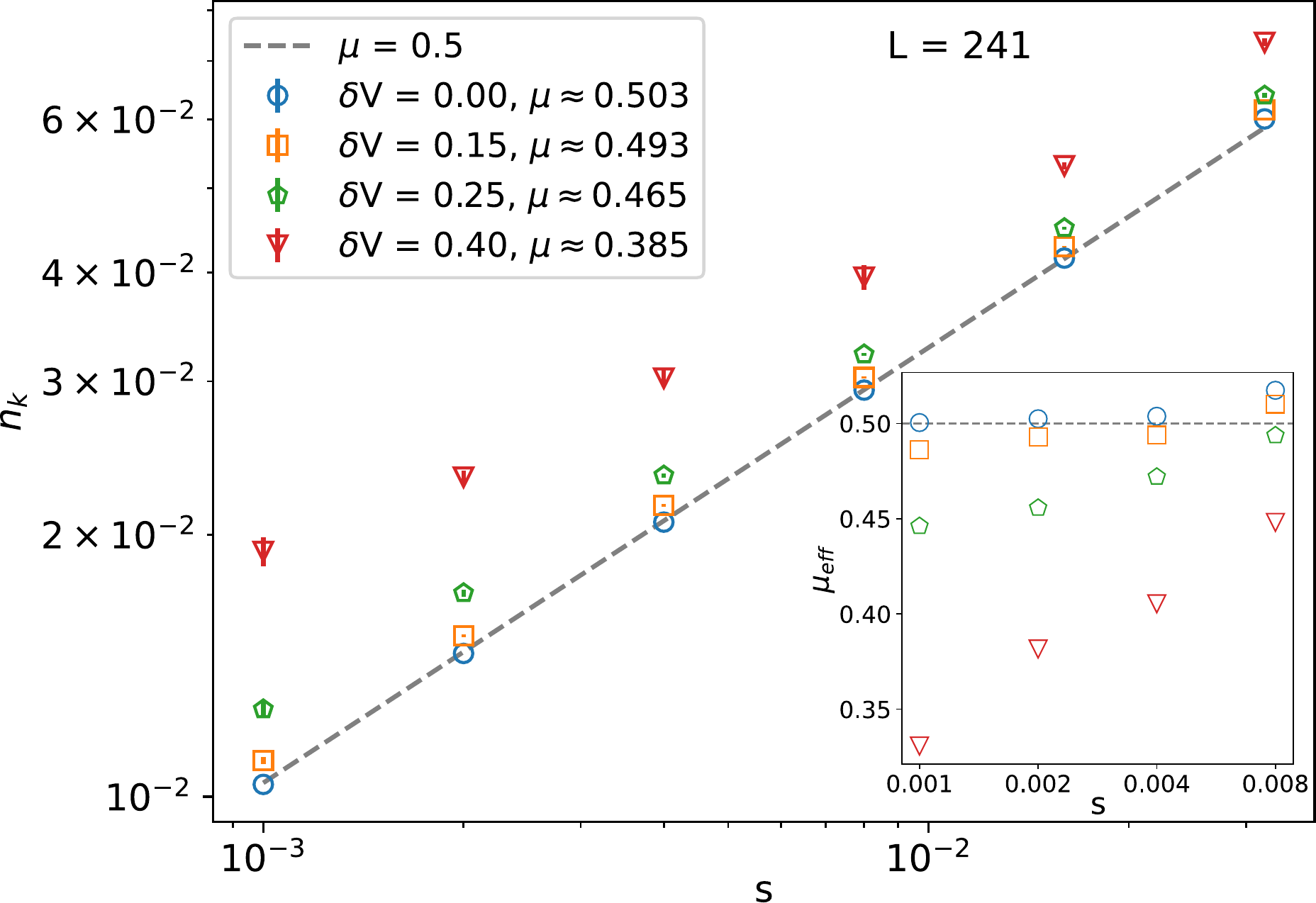}
    \caption{Density of kinks $n_k$ as a function of the sweep rate $s$ for disorder strengths $\delta V = 0.15, 0.25, 0.4$, for a system size  $L = 241$. The clean case (blue circles) is shown for reference. The dashed gray line indicates the theoretical prediction for the clean Ising transition. As disorder strength increases, the scaling of $n_k$ progressively deviates from the clean power-law scaling. We averaged over up to 250-330 independent samples depending on the strength of the disorder. Inset: effective exponents $\mu_{\textnormal{eff}}$ extracted by fitting the slopes of sets of three consecutive points in the main panel. In the disordered case, $\mu_{\textnormal{eff}}$ decreases progressively for slow sweep rates.}
    \label{fig:kzl241}
\end{figure}

{\bf Deviation from the power-law scaling.} First, we demonstrate that disorder has a trackable  impact on the critical Kibble-Zurek scaling. Fig.~\ref{fig:kzl241} shows the scaling of the density of kinks $n_k$ with the sweep rate $s$ for various values of disorder strength $\delta V$ and for a fixed system size $L = 241$. We present our results in a log-log scale, where the critical scaling in the clean case is linear (dashed line). As the disorder $\delta V$ increases, the slope of the scaling $\mu_\mathrm{eff}$ decreases, deviating markedly from the clean power-law behavior. This deviation becomes even more evident in the inset, where we track an effective exponent $\mu_{\text{eff}}$ from sets of three consecutive data points in Fig.~\ref{fig:kzl241}. Plotted against the first (leftmost) point in each set, $\mu_{\text{eff}}$ decreases with decreasing sweep rate $s$. This trend becomes more pronounced as $\delta V$ increases, leading to a systematic deviation of the measured exponent $\mu$ from the theoretical predictions for the Ising transition.

\begin{figure}[h]
    \centering
    \includegraphics[width=0.45\textwidth]{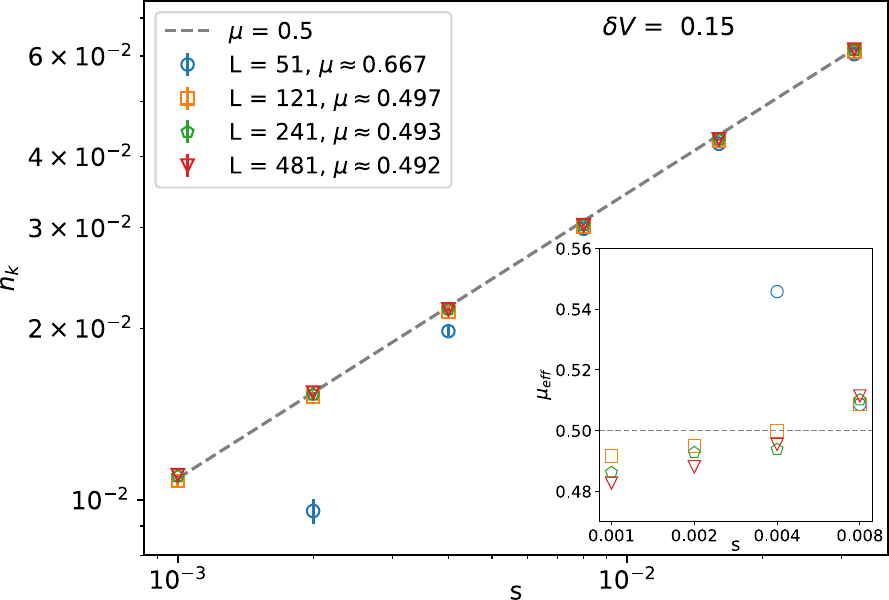}
    \caption{Scaling of the kink density $n_k$ with sweep rate $s$ for disorder strength $\delta V = 0.15$ and various system sizes. Results were averaged over up to 300 individual samples. Kibble-Zurek exponents obtained by fitting all shown points are listed in the legend. Inset: effective critical exponent $\mu_{\textnormal{eff}}$ extracted from sets of three consecutive points of the main panel. For slower sweep rates and larger system sizes the effect of disorder becomes more pronounced.}
    \label{fig:kzdis0.15}
\end{figure}

{\bf The crossover.} In Fig.~\ref{fig:kzdis0.15} we present the scaling of the density of kinks $n_k$ versus the sweep rate $s$ for several system sizes at fixed disorder strength $\delta V = 0.15$. By contrast to the clean case, $\mu$ decreases with system size, resulting in a stronger deviation from the clean Ising prediction. 
Such crossover between clean and disordered physics as a function of the system size is a documented feature in equilibrium~\cite{PhysRevB.70.054430,Laflorencie2022,chepiga2024resilient}.
The inset shows $\mu_{\text{eff}}$, extracted using the same method as in Fig.~\ref{fig:kzl241}, now plotted for each system size. As before, $\mu_{\text{eff}}$ decreases with decreasing $s$, but the deviation is more pronounced for faster sweep rates and larger system sizes. A notable exception is the case $L = 51$, where $\mu_{\text{eff}}$ increases instead. This is not unexpected: for small systems and sufficiently slow quenches, the finite energy gap at the critical point allows for an adiabatic evolution throughout the quench~\cite{zurek2005dynamics}. This may explain why the effect of disorder was not observed in recent experiments probing Ising transition with 51 Rydberg atoms~\cite{keesling2019quantum}.

\begin{figure}[h]
    \centering
    \includegraphics[width=0.45\textwidth]{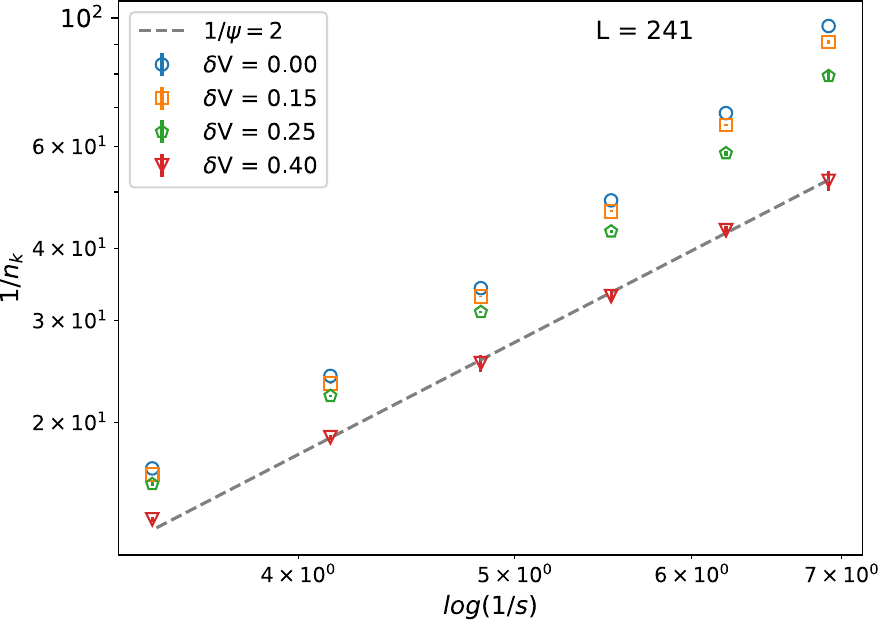}
    \caption{Scaling of the inverse of the density of kinks $n_k$ with the logarithm of the inverse of the sweep rate $s^{-1}$ for various disorder strengths on a log--log plot. The dashed gray line indicates the theoretical prediction for infinite-randomness criticality, $1/n_k \sim \log^{1/\psi}(s^{-1})$ with $1/\psi = 2$~\cite{fisher1995critical}. Results were obtained for a chain with $L = 241$ sites. As the disorder strength increases, the numerical results increasingly approach the predicted infinite-randomness scaling behavior.}
    \label{fig:logsquaredl241}
\end{figure}

{\bf Dynamical signatures of the infinite randomness.} 
Previous studies have shown that in the infinite randomness universality class, the scaling of the density of kinks $n_k$ with the sweep rate $s$ follows a much weaker, logarithmic form, specifically $n_k \sim [\log(s^{-1})]^{-2}$, rather than a power law typical for the clean transitions~\cite{dziarmaga2006dynamics, caneva2007adiabatic}. To test how well the KZ mechanism can capture infinite-randomness criticality on experimentally accessible system sizes, we plot $1/n_k$ versus $\log(1/s)$ on a log-log scale in Fig.~\ref{fig:logsquaredl241}. The dashed line corresponds to the analytical prediction. As $\delta V$ increases, the data increasingly conforms to this behavior, with the case $\delta V = 0.4$ displaying an almost perfect match.

{\bf Localization  of the floating phase.} 
We study the effect of disorder on the equilibrium floating phase using density matrix renormalization group (DMRG) simulations. In contrast to the Ising transition, whose critical point can be shifted by disorder, Rydberg arrays exhibit extended regions of floating phases~\cite{maceira2022conformal, zhang2025probing}. Here, we focus on a point deep inside one such region, between the period-4 and period-5 ordered lobes. We characterize the floating phase through three hallmarks: quasi-long-range incommensurate correlations, the structure factor, and the scaling of entanglement entropy.

\begin{figure}[t!]
    \centering
    \includegraphics[width=\linewidth]{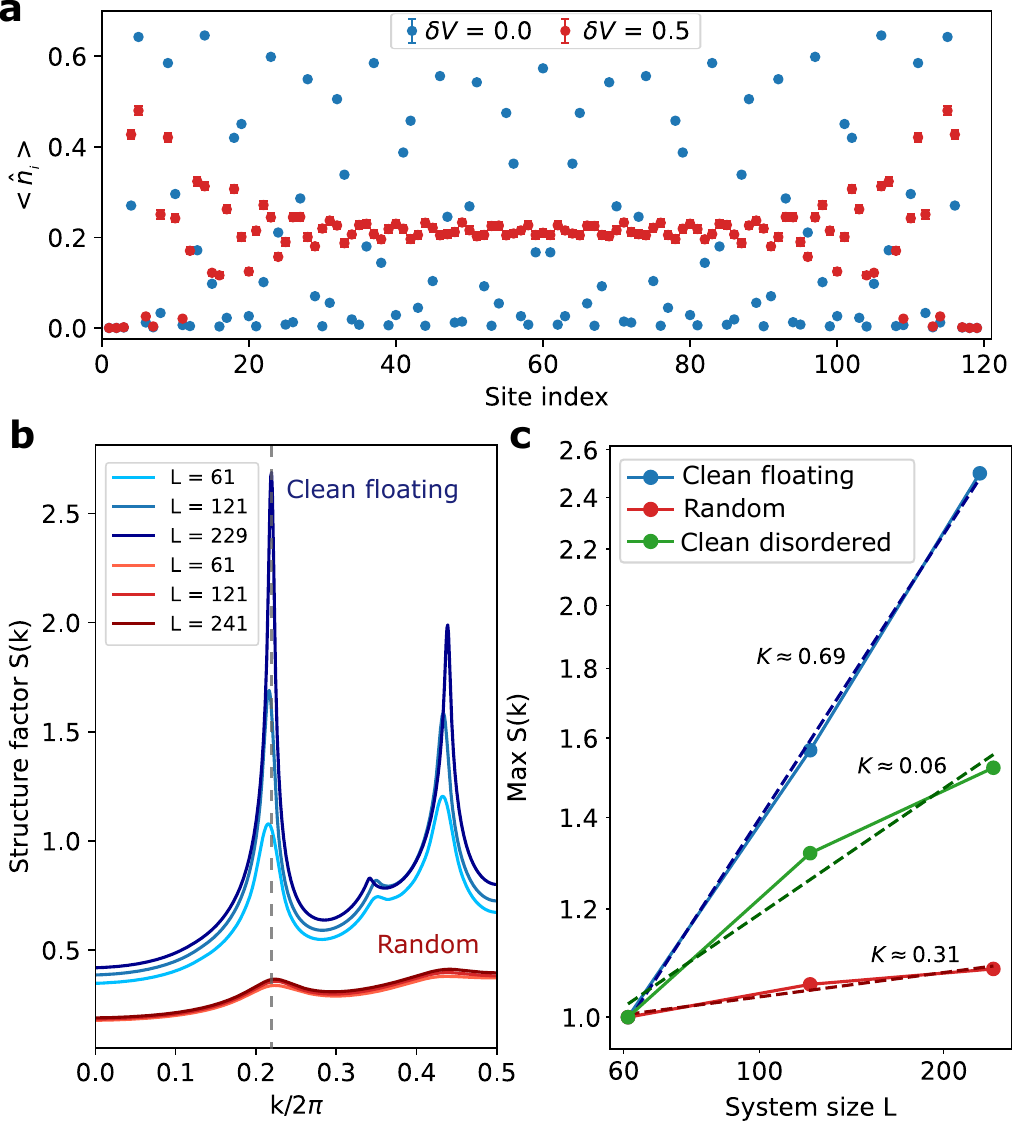}
   \caption{Numerical evidence of localization of the floating phase. (a) Local density profile of a chain of Rydberg atoms for a disorder strength $\delta V = 0.5$ (orange) and for the clean case (blue) deep inside the floating phase for $L = 121$. In the disordered case, obtained by averaging over $\sim1700$ disorder configurations, the edge effects are localized within about 30 lattice sites. We present the results for $\Omega = 1$, $\Delta = 3.5$, and $R_b/a = 4$, deep inside the floating phase. (b) Structure factor $S(k)$ for the clean (blue) and disordered (red) systems at the specified point. (c) Scaling of the first peak height of the structure factor shown in (b), which defines the Luttinger liquid exponent $K$. For reference, we also include results for the clean case inside the disordered phase, exhibiting incommensurate short-range correlations at $\Delta = 2.5$ (green).}
  \label{fig:friedel}
\end{figure}

\textit{Friedel oscillations}. There are known examples in the literature of commensurate Luttinger liquids with repulsive interactions being localized by arbitrary weak disorder~\cite{giamarchi2003quantum, berkovits2012entanglement}. To probe this phenomenon in our setting, we analyze Friedel oscillations in the local density of Rydberg excitations, which reveal the system’s ability (or failure) to heal and screen impurities, here represented by the chain edges. In the critical floating phase, Friedel oscillations follow the predictions of boundary conformal field theory, exhibiting a slowly decaying profile, as illustrated for the clean case in Fig.~\ref{fig:friedel}(a) (blue curve). By contrast, introducing disorder typical for this regime, $\delta V = 0.5$ (see End Matter for details), dramatically alters the oscillations: edge effects are almost entirely suppressed beyond $\sim 30$ sites, indicating a finite localization length.

\textit{Structure factor}. An alternative approach that gives direct access to the incommensurability, both numerically and experimentally~\cite{zhang2025probing, soto2025numerical}, is to analyze the location of the first peak in the static structure factor: 
\begin{equation}
S(q) = \frac{1}{L}\sum_{i,j} e^{iq(i-j)} C_{ij},
\end{equation}
with $
C_{ij} = \langle \hat{n}_i \hat{n}_j \rangle - \langle \hat{n}_i \rangle \langle \hat{n}_j \rangle $ being the density-density correlation function. In Fig.~\ref{fig:friedel}(b), we compare the structure factor for clean and disordered systems for several system sizes. Remarkably, incommensurability is not destroyed by disorder, a phenomenon that has recently been reported in the context of Majorana wires~\cite{chepiga2024resilient}. Surprisingly, the leading wave-vector in the clean and disordered case deviates by no more than $O(10^{-3})$.

The main qualitative difference between the two cases is the height of the peak, which is directly related to the correlation length: large and rapidly growing with the system size inside the clean floating phase~\footnote{In a Luttinger liquid the peak of the structure factor follows the scaling relation $S(L) \sim L^{K}$, where $K$ is the Luttinger liquid exponent \cite{sengupta2002bond}}, and small and stable in the presence of disorder (see Fig.\ref{fig:friedel}(c) for a quantitative comparison). For comparison we also include the results for a gapped disordered phase in the clean case.

\begin{figure}[h!]
    \centering
    \includegraphics[width=0.48\textwidth]{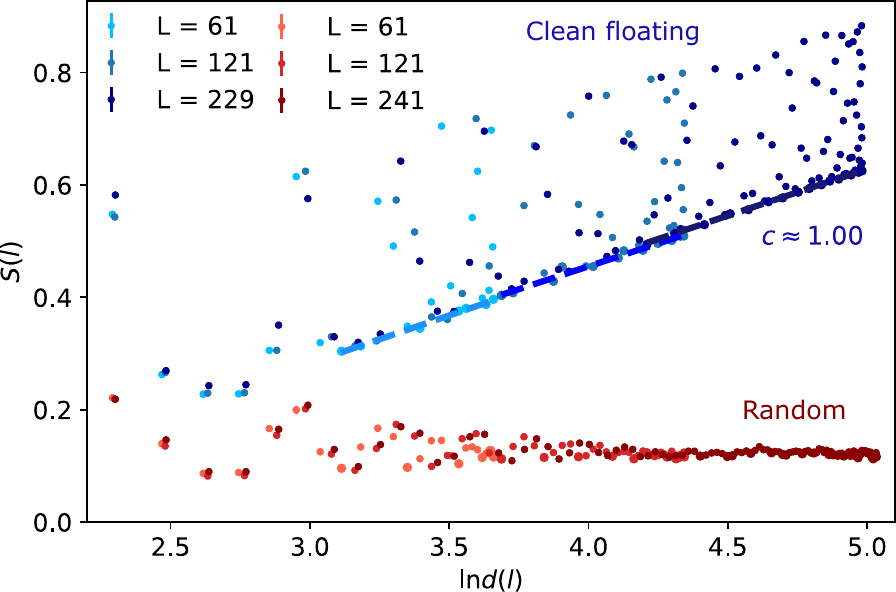}
    \caption{Scaling of the entanglement entropy $S$ as a function of the conformal distance $d(l)$ for three system sizes for a clean system ($\delta V = 0.0$) and a disordered system with $\delta V = 0.5$ deep inside the floating phase. The clean system shows a semilog scaling of the entanglement entropy with the conformal distance, consistent with $c = 1$, indicating a floating phase, while the disordered case saturates to a constant value, indicating a gapped phase. The specific point in the phase diagram is $\Omega = 1$, $\Delta = 3.5$,  $R_b/a = 4$.}
    \label{fig:centralcharge}
\end{figure}

\textit{Entanglement entropy}. The localization of the floating phase is also reflected in the behavior of the entanglement entropy, as shown in Fig.~\ref{fig:centralcharge}. In the clean system, the floating phase is an incommensurate Luttinger liquid. According to conformal field theory~\cite{calabrese2009entanglement,capponi2013quantum}, the entanglement entropy of a block of size $l$ in a finite chain with open boundary conditions grows as
\begin{equation}
S(l) = \frac{c}{6}\ln d(l) + s_1 + \ln g, 
\end{equation}
where $c=1$ is the central charge of the Luttinger liquid, $d(l)=\frac{2L}{\pi}\sin\!\left(\frac{\pi l}{L}\right)$ is the conformal distance, $\ln g$ states for boundary entropy, and $s_1$  is a nonuniversal constant. 
Our clean data presented in Fig.~\ref{fig:centralcharge} show precisely this logarithmic growth fully consistent with the critical nature of the floating phase. 

By contrast, in the presence of quenched disorder the entanglement entropy saturates to a finite value. In one dimension, such saturation is characteristic of low-energy states of gapped Hamiltonians, which obey an area law for entanglement entropy \cite{calabrese2004entanglement, hastings2007area, eisert2010colloquium}. Consequently, the bipartite entanglement entropy does not grow with subsystem size but instead levels off once the subsystem exceeds the localization length \cite{berkovits2012entanglement}. The saturation of the entanglement $S(\ell)$ therefore provides a direct and unambiguous signature of the disorder-induced localization of the floating phase.

{\bf Discussion. }
We have demonstrated that disorder in the interaction potential caused by the finite width of optical tweezers has a significant impact on the measurable Kibble-Zurek (KZ) critical exponent $\mu$.  This provides the first consistent explanation for the systematic overestimation of the KZ exponent in numerical simulations compared to experimental data~\cite{keesling2019quantum}. We show that deviations from the clean physics will be even more visible on larger system sizes, slower sweep rates and for smaller interatomic distances. 
Our results highlight practical constraints on scaling Rydberg-based quantum simulators, as accessing comparable regimes in longer chains will require either improved trapping or higher Rydberg states to maintain sufficient interaction strength while increasing the interatomic distances.

We expect our results to underestimate the actual impact of disorder. First, for numerical efficiency we introduced disorder only through the coupling constant $V_i$, thereby underestimating lattice distortions by relying on a first-order error propagation. Second, in experiments, the trapping potential is typically switched off before measurement, which further enhances the distortion.

We perform numerical simulations with long-range interactions and, remarkably, the results shown in Fig.~\ref{fig:logsquaredl241} are consistent with Fisher’s infinite-randomness fixed point of the random Ising model with only nearest-neighbor interactions. In this regard, our findings differ from the predictions of strong-disorder renormalization group studies of the ferromagnetic Ising chain with random long-range interactions and random fields~\cite{juhasz2014random}, which predict a transition characterized by a finite dynamical critical exponent $z$ that depends on the decay exponent of the interaction strength. We hope this work will stimulate further theoretical progress in understanding the role of long-range couplings in disordered systems.

Furthermore, we have shown that the floating phase is highly sensitive to disorder and exhibits striking signatures of localization, including saturated entanglement entropy and a finite correlation length. At the same time, and quite unexpectedly, we find that the dominant wave vector $q$ remains unchanged. This observation raises several fundamental questions: Is there a disorder version of the $Z_3$- and $Z_4$-chiral transitions~\cite{huse1982domain, huse1984commensurate, chepiga2019floating, chepiga2021kibble}? Do commensurate–incommensurate transitions~\cite{bak1979phase, balatsky1984commensurate, tsvelik1992influence} persist in the presence of disorder, and with what characteristics? And if the floating phase is indeed localized, how is the Pokrovsky–Talapov transition~\cite{pokrovsky1979ground}, which in the clean system separates the ordered and floating phases, modified?

{\bf Acknowledgments.}
We thank Hannes Bernien and Nicolas Laflorencie for
insightful discussions. This research was supported
by Delft Technology Fellowship. Numerical simulations were performed at the DelftBlue HPC and at
the Dutch national e-infrastructure with the support of the SURF Cooperative.

\pagebreak

\section{End Matter}
\textbf{Estimation of the disorder strength}.
Our analysis is based on the experimental setup of Ref.~\cite{bernien2017probing} and its follow-up work~\cite{keesling2019quantum}. In these experiments, the authors work with $^{87}R_b$ atoms trapped in optical tweezers with a wavelength of $808$ nm, a beam waist of $w_0 \approx 0.9\, \mu$m, and a trap depth of about $U_0 \approx k_B\times 1$ mK. The atoms are cooled to a temperature of $T \approx 12\,\mu$K, which leads to positional fluctuations of approximately $\delta x \approx 0.12\,\mu\text{m}$ in the radial direction and $\delta y \approx 0.6\,\mu\text{m}$ along the longitudinal axis~\cite{bernien2017probing}. Such uncertainties can significantly affect the interaction strength between atoms.  This is not the only source of error, as it is well documented in Ref.~\cite{bernien2017probing}. However, as far as we understand, the uncertainty in the location of the atoms is the only relevant source of disorder that alters the nature of the transition, while other imperfections can only add marginal contributions, making the results only worse, but not healing the system back to the clean case. 

To quantify this effect, we apply linear error propagation. For a general function $f(x,y)$ with uncertainties $\delta x$ and $\delta y$, the propagated uncertainty is
\begin{equation}
\delta f(x,y) = \sqrt{\left( \frac{\partial f}{\partial x} \right)^2 (\delta x)^2 + \left( \frac{\partial f}{\partial y} \right)^2 (\delta y)^2 }.
\end{equation}

In our case, the relevant dependence arises from the interatomic spacing $a$ entering the van der Waals interaction:
\begin{equation}
    V = \frac{V_0}{a^6}\sum_{i<j}\frac{1}{|i-j|^6}\hat n_i \hat n_j.
\end{equation}
Propagating the uncertainty in $a$ gives
\begin{equation}
    \delta V = 6\frac{V_0}{a^7}\,\delta a \sum_{i<j}\frac{1}{|i-j|^6}\hat n_i \hat n_j 
    \Rightarrow \frac{\delta V}{V} = 6\frac{\delta a}{a}.
\end{equation}

To leading order, the uncertainty in the relative displacement is
\begin{equation}
    \delta a = \Delta x = \sqrt{2}\,\delta x \approx 0.12\sqrt{2}\,\mu\text{m}.
\end{equation}
This estimate underestimates the true uncertainty in $V(r)$, as it neglects longitudinal fluctuations of the interatomic distance when evaluated at $y=0$. Nonetheless, we adopt it as a conservative lower bound.

For completeness, including second-order perturbative corrections would yield
\begin{equation}
    \delta a = \sqrt{ (\Delta x)^2 + \frac{(\Delta y)^4}{2x^2} }.
\end{equation}

Finally, we neglect fluctuations in the Rabi frequency $\Omega$, which would further increase the total uncertainty. Hence, the values reported here should be regarded as conservative lower bounds on the disorder strength in experiments.

Typical interatomic distances observed in experiments are~\cite{bernien2017probing}:
\begin{itemize}
\item $a \approx 8.7$–$4.3\,\mu\text{m}$ for the period-2 phase,
\item $a \approx 4.3$–$2.9\,\mu\text{m}$ for the period-3 phase,
\item $a \approx 2.9$–$2.2\,\mu\text{m}$ for the period-4 phase.
\end{itemize}

Based on these ranges, we select four representative disorder strengths based on first order error propagation corresponding to characteristic points of the phase diagram:
\begin{itemize}
\item Ising transition ($a \approx 5.74\,\mu\text{m}$): $\delta V \approx 0.17\, V$.
\item Potts point in the period-3 phase \cite{maceira2022conformal} ($a \approx 3.95\,\mu\text{m}$): $\delta V \approx 0.26\, V$.
\item Ashkin–Teller point in the period-4 phase \cite{maceira2022conformal} ($a \approx 2.7\,\mu\text{m}$): $\delta V \approx 0.38\, V$.
\item Floating phase between periods 4 and 5: ($a \approx 2.17\,\mu\text{m}$): $\delta V \approx 0.47\, V$.
\end{itemize}

In our simulations, we therefore consider four levels of disorder reflecting these estimates: $\delta V = 0.15$, $\delta V = 0.25$, $\delta V = 0.40$, and $\delta V = 0.50$.

As a final remark, the van der Waals interaction is intrinsically \textit{asymmetric} with respect to atomic displacements. For example, for an interatomic spacing of $a = 2.17\,\mu\mathrm{m}$---the value used in our study of the floating phase---the interaction strength varies between $\delta V_-/V \approx 0.36$ and $\delta V_+/V \approx 0.67$. The effective half-width, $(\delta V_- + \delta V_+)/2V \approx 0.5$, coincides with our estimate, but the distribution is skewed toward stronger interactions, resulting in a shifted mean. This mean shift primarily leads to a small displacement of the critical point in parameter space, without altering the universal critical behavior.
  
\textbf{Ground state calculations}.
The microscopic model of Rydberg atoms  was simulated using two-sites Density Matrix Renormalization Group (DMRG)~\cite{white1992density, schollwock2011density} in the matrix product state (MPS) formalism~\cite{schollwock2011density}. The many-body system Hamiltonian was expressed as a matrix product operator (MPO)~\cite{verstraete2004matrix}, and the algebraically decaying 2-sites interactions were approximated by a sum of 6 exponentials~\cite{pirvu2010matrix, schollwock2011density} $1/r^6 = \sum_{i=1}^{6} c_i\lambda_i^r$, where $c_i$ and $\lambda_i$ are parameters determined by minimizing the cost function, defined as:
\begin{equation}
\sum_{r=1}^L\abs{\frac{1}{r^6}- \sum_{i=1}^{6}c_i\lambda_i^r}.
\end{equation}
This approximation provides an almost perfect fitting for the interaction of $\sim20$ atoms distance and a cost function error of $\sim3.5\times 10^{-18}$ for $L = 241$.  Ground state convergence was assumed when the variation in the total energy after two consecutive sweeps $<10^{-12}$. 

\textbf{Simulation of dynamics}. Time evolution was performed using a 2-sites time-dependent variational algorithm (TDVP)~\cite{haegeman2011time, haegeman2016unifying, paeckel2019time}.  In this case, the algebraically decaying 2-sites interactions were approximated by a sum of 4 exponentials. This approximation provides almost perfect fitting for $\sim 10$ atoms.  For each time-step, convergence of the Lanczos algorithm was assumed when the difference on the energy after two consecutive iterations was $< 10^{-7}$.

\textbf{Kibble-Zurek mechanism protocol}. The starting point for the Kibble-Zurek (KZ) mechanism was set at a distance of $15 \times \sqrt{s}$ normalized by the interatomic strength constant $V_0$ to ensure the system started in the adiabatic regime. This choice is motivated because for an Ising transition, adiabaticity breaks down when the distance to the critical point reaches $|\varepsilon| \sim \sqrt{s}$. 
The van der Waals interaction strength for each site was held fixed $V_i = 1 \pm \delta V_i$ and $\Delta$ and $\Omega$ were swept around the phase transition in a direction perpendicular to the critical line identified in the phase diagram of Fig.1 (right) in Ref.~\cite{rader2019floating}. The quench was terminated at $\Omega = 0$.

The number of kinks was quantified as the density of domain walls, following the definition proposed in Ref.~\cite{soto2024resolving}. Specifically, for a period-2 structure, two types of domains—A and B—can be distinguished based on whether the Rydberg excitation occupies odd or even lattice sites. A kink is then defined as a domain wall where the excitation pattern switches between domains A and B. Comparable results were obtained using the \textit{non-order parameter} introduced in the same reference. In this alternative approach, a kink is identified through the local pattern $(1 - \fullcircle\emptycircle -\emptycircle\fullcircle)$. While both methods yield similar results, notable differences exist. For example, the configuration $\fullcircle\emptycircle\emptycircle\emptycircle\fullcircle$ is not counted as a kink using the domain-wall method but is counted as two kinks under the non-order parameter definition. For a more detailed discussion of kink definitions and their influence on KZ scaling (see Ref.~\cite{garcia2024quantum}). All quench protocols crossed the critical point at $(\Omega/V_0, \Delta/V_0) \approx (0.494, 1)$, corresponding to the estimated transition point of the clean model~\cite{rader2019floating}.

For each sweep rate, the slow quench process was repeated between 20 and 300 times, depending on the specific sweep rate and system size, and the resulting kink densities were averaged. The uncertainty in the mean was estimated as $1.96 \times$ the standard error, corresponding to a 95\% confidence interval for the estimated mean.

\begin{figure}[h!]
    \centering
    \includegraphics[width=\linewidth]{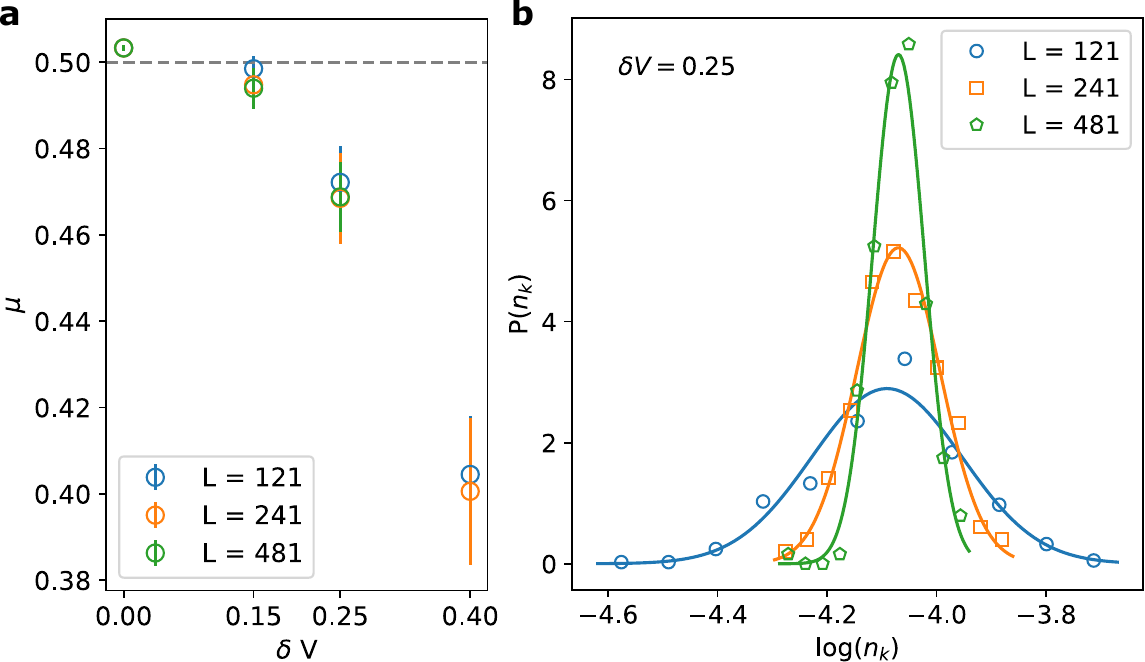}
    \caption{Additional results on finite-size effects in  Kibble-Zurek scaling. (a) The critical exponent $\mu_\mathrm{eff}$ is shown as a function of disorder strength $\delta V$ for various system sizes. The effect of disorder slowly increases with system size, but due to large error bars it would be challenging to quantify the effect. (b) Probability distribution of kink density after a slow quench. We show the distribution of $\log(n_k)$ for  $L=121, 241, 481$  averaged over 430, 250, and 200 independent samples correspondingly. Histograms were computed using 11 fixed bins. As the system size increases, the distribution narrows, reflecting reduced fluctuations in larger systems. Solid lines correspond to a Gaussian fit.}
    \label{fig:mudis}
\end{figure}

\textbf{Finite-size effects of Kibble-Zurek mechanism in the presence of disorder.} We further discuss finite-size effects in the deviation from the clean Ising scaling in KZ dynamics. In Fig.~\ref{fig:mudis}(a) we display the extracted values of $\mu$ as a function of $\delta V$ for various system sizes. The results confirm that the measureble KZ exponent $\mu_\mathrm{eff}$ decreases with increasing disorder and also with system size. Additionally, the error bars on the fitted exponents grow with $\delta V$, reflecting a growing deviation from simple power-law scaling.

To probe the dynamical fingerprints of the infinite–randomness critical point, we investigate how the kink–density distribution evolves with system size at a fixed sweep rate. As illustrated in Fig.~\ref{fig:mudis}(b), the histogram of \(\log n_k\) remains centered on a clear average value, while its overall spread becomes progressively narrower for larger chains. A qualitatively similar trend was reported for disordered quantum Ising chains driven through the transition in the KZ regime (see Ref.~\cite{caneva2007adiabatic}). The observed reduction of sample-to-sample fluctuations reflects the increasing number of effectively independent domains that form once the dynamics freezes out, yielding a post-quench observable that is self-averaging even though equilibrium quantities at the underlying infinite-randomness fixed point itself remain non-self-averaging.

\newpage

\bibliographystyle{apsrev4-2}
\bibliography{bibliography}

\end{document}